\documentclass{SCIS2022}
\newcommand{\mb}[1]{{  \mathbf  #1}}  
\begin{document}
\ArticleType{RESEARCH PAPER}
\Year{2022}
\Month{}
\Vol{}
\No{}
\DOI{}
\ArtNo{}
\ReceiveDate{}
\ReviseDate{}
\AcceptDate{}
\OnlineDate{}

\title{Near-Field Wideband Channel Estimation for Extremely Large-Scale MIMO}{Near-Field Wideband Channel Estimation for Extremely Large-Scale MIMO}

\author[]{Mingyao Cui}{}
\author[]{Linglong Dai}{{daill@tsinghua.edu.cn}}


\AuthorCitation{Mingyao Cui and Linglong Dai}


\address[]{ Department of Electronic Engineering, Tsinghua University, \\ 
Beijing National Research Center for Information Science and Technology (BNRist), Beijing {\rm 100084}, China}

\abstract{
Extremely large-scale multiple-input-multiple-output (XL-MIMO) at millimeter-wave (mmWave) and terahertz (THz) bands plays an important role in supporting extreme high beamforming gain as well as ultra-wideband spectrum resources. Unfortunately, accurate wideband XL-MIMO channel estimation suffers from the new challenge called as the near-field beam split effect. Prior works either neglect the accurate near-field channel model or fail to exploit the beam split effect, resulting in poor channel estimation accuracy for wideband XL-MIMO. To tackle this problem, this paper proposes a bilinear pattern detection (BPD) based approach to accurately recover the wideband XL-MIMO channel. {\color{black}Specifically, by analyzing the characteristics of near-field wideband channels, we first reveal the bilinear pattern of the near-field beam split effect, which implies that the sparse support set of near-field channels in both the angle and the distance domains can be regarded as a linear function against frequency. Then, inspired by the classical simultaneously orthogonal matching pursuit technique, we use the bilinear pattern to estimate the angle-of-arrival (AoA) and distance parameters of each near-field path component at all frequencies. In this way, the entire wideband XL-MIMO channel can be recovered by compressed sensing algorithms. Moreover, we provide the computational complexity of the proposed algorithm compared with existing algorithms. }Finally, simulation results demonstrate that our scheme can achieve the accurate estimation of the near-field wideband XL-MIMO channel in the presence of near-field beam split effect.
}


\maketitle


\section{Introduction}
 
Compared with the existing massive multiple-input-multiple-output (MIMO) with dozens or hundreds of antennas for 5G, extremely large-scale MIMO (XL-MIMO) embraces a 10-fold increase in the scale of antenna array, which enables the considerable improvement of the spatial multiplexing gain~\cite{6Gchallenge_Rappaport2019,6GVision_Zhang2019, UMIMO_Han21,THzCom_Lin2016}. Moreover, benefiting from the high beamforming gain achieved by extremely narrow pencil-like beams, XL-MIMO is essential to alleviate the serious path loss of millimeter wave (mmWave) and terahertz (THz) signals with abundant spectrum resources~\cite{mmWave_Akdeniz2014,THzdistance_Akyildiz2018,THzsurvey_Akyildiz2014}. Consequently, the natural integration of high-frequency wideband communications and XL-MIMO is regarded as a promising technology in future 6G networks. 

To unleash the performance superiority of wideband XL-MIMO, it is essential to acquire accurate channel state information (CSI). Unfortunately, due to the enormous number of antennas, XL-MIMO channel estimation requires an unacceptable pilot overhead, particularly when the base station (BS) is deploying the hybrid analog and digital precoding architecture with only a small number of radio-frequency (RF) chains~\cite{precoding_Sohrabi2016,Overview_Heath16}. Thereby, how to exploit the channel structure to reduce the pilot overhead for channel estimation has always been a hot topic from 5G massive MIMO to 6G XL-MIMO.
\subsection{Prior Works}
There are extensive works to explore the sparse structure of mmWave and THz channels to design channel estimation algorithms with low pilot overhead. These works can be generally categorized into three types: far-field narrowband~\cite{CE_OMP_Lee16,CE_SMP_Huang19,SWOMP_Robert18,CE_SOMP_Gao16,CS_Eldar2015}, far-field wideband~\cite{THzBS_Gao2021,BSD_Tan21,THzCE_Dov2021}, and near-field narrowband~\cite{NearCE_Yang2021,NearCE_Han2021,NearCE_Cui2022} channel estimations.

The first type of \emph{far-field narrowband} channel estimation schemes mainly works for 5G massive MIMO, and they usually assume that the bandwidth is limited, e.g., around several hundreds of MHz. The array aperture of massive MIMO is not large enough, resulting in the dominance of far-field transmission environments, where a channel path can be modeled by the planar wave model. In this case, the multipath far-field channel is built up by the superposition of several planar waves~\cite{Overview_Heath16}. As each planar wavefront is dependent on a certain angle-of-arrival (AoA), and the number of channel paths is usually small, this far-field channel exhibits sparsity in the angle domain. Thus it can be recovered by compressed sensing (CS) based algorithms with low pilot overhead. For instance, \cite{CE_OMP_Lee16} utilized orthogonal matching pursuit (OMP) scheme to estimate the angle-domain sparse channel at a single frequency. Moreover, several works have considered to improve the OMP algorithm by using the channel sparse structure at multiple frequencies.
Specifically, due to the limited bandwidth, the common sparse support assumption has been widely used to transform the multi-frequency channel estimation into a multiple measurement vector (MMV) problem, where the sparse support sets are the same at different frequencies.
Then, various MMV-specific algorithms can be utilized to estimate the far-field channel, such as the simultaneously OMP (SOMP)~\cite{SWOMP_Robert18,CE_SOMP_Gao16} and the group-sparse Bayesian CS approach~\cite{CS_Eldar2015}.

Compared with the \emph{far-field narrowband} schemes above, the \emph{far-field wideband} channel estimation algorithms assume a pretty wide bandwidth. In this condition, although communications are still operating in far-field environments, the widely-used common sparse support assumption~\cite{SWOMP_Robert18,CE_SOMP_Gao16}  does not hold anymore due to the beam split effect~\cite{THzBS_Gao2021,BSD_Tan21}. The beam split effect indicates that owing to the wide bandwidth, the spatial channel directions are separated from each other at different frequencies, which results in a frequency-dependent angle-domain sparse structure and severely degrades the performance of classical MMV-specific solutions. Recently, a few advanced signal processing techniques have been proposed to cope with this beam split effect in far-field scenarios. Specifically, \cite{BSD_Tan21} proved the beam split pattern in the angle domain, i.e., the sparse support set of a physical angle grows linearly with the subcarrier frequency. Then, a beam split pattern detection (BSPD) based algorithm was proposed in~\cite{BSD_Tan21} to construct the match between a certain physical angle and the beam split pattern for recovering the far-field wideband channel. Besides, in~\cite{THzCE_Dov2021}, the authors proposed to construct a series of dictionary matrices for each subcarrier frequency to match the beam split pattern, based on which a generalized simultaneous OMP (GSOMP) was applied to overcome the beam split effect in the far field.

For the third category of the \emph{near-field narrowband} algorithms, they place heavy emphasis on the fundamental change from 5G massive MIMO to 6G XL-MIMO by considering the near-field propagation. The deployment of XL-MIMO makes the near-field propagation highly possible to happen, particularly for high-frequency communications~\cite{NearMag_Cui2022}. To be specific, the radius of near-field areas, which is also called as Fraunhofer distance, is proportional to the square of array aperture and frequency~\cite{fresnel_Selvan2017}. The Frauhofer distance may reach several dozens or hundreds of meters in high-frequency XL-MIMO systems. As opposed to the far-field channel model based on the planar wavefront assumption, the multi-path near-field channel model is built up by the superposition of multiple \emph{spherical waves}~\cite{NearBF_Zhang2022}. Although the number of channel paths is still limited, the angle-domain near-field channel is not sparse any more, since each spherical wave is dependent on a range of physical angles~\cite{NearCE_Cui2022,NearCE_Yang2021}. Thus, existing angle-domain sparsity based techniques cannot work well in near-field environments. To address this problem, \cite{NearCE_Cui2022} proposed to transform the near-field channel into the joint angle-distance domain (polar domain). As an angle-distance sample contains enough information to describe a spherical wave, the near-field channel can be then sparsely represented in the polar domain. Accordingly, the polar-domain SOMP algorithms were used in \cite{NearCE_Cui2022} to recover the near-field channel accurately. 


Despite the extensive studies on XL-MIMO channel estimation, the above three categories of techniques neglect the crucial topic of \emph{near-field wideband} channel estimation. In other words, they overlook the important integration of the near-field and beam split effect. For wideband XL-MIMO communications, both the array aperture and bandwidth are pretty large.  Such a wideband XL-MIMO system poses a \emph{near-field beam split} effect~\cite{NFBF_Cui2021}, where the sparse channel support set in the polar domain differs from each other at different frequencies. Furthermore, as all prior techniques (far-field narrowband, far-field wideband, and near-field narrowband approaches) mismatch this polar-domain frequency-dependent sparse structure, they will suffer from a serious performance loss in wideband XL-MIMO systems. To the best of our knowledge, the essential near-field wideband channel estimation has not been studied in the literature.

%

\subsection{Our Contributions}
To fill in this gap, we propose a bilinear pattern detection (BPD) based algorithm to realize near-field wideband channel estimation by exploring the unique polar-domain sparse structure with near-field beam split. Our contributions are summarized as follows. 
\begin{itemize}
\item First, we comprehensively analyze the polar-domain channel sparse structure over different subcarrier frequencies when the near-field beam split effect is considered.
By analyzing the mutual coherence of near-field array response vectors at different frequencies, we reveal the phenomenon that both the angle-domain and distance-domain sparse support sets grow linearly with the subcarrier frequency, which is defined as the bilinear pattern (BP) of near-field beam split effect. This phenomenon indicates that the physical location of a scatter (or user) in a channel path has a one-to-one matching relationship with a certain pattern of the frequency-dependent sparse support sets in the polar domain. This relationship can be used to accurately estimate the physical location of each path. 
\item Then, a BPD based channel estimation algorithm is proposed according to the revealed phenomenon above. We first construct the polar-domain bilinear patterns for all angle-distance samples. Then, for each path component, we use these bilinear patterns to accumulate the power of each angle-distance sample from the entire bandwidth and capture the sample with the largest power as the estimated physical location. 
We carry out the above procedure several times to successively detect all channel paths. 
As the proposed algorithm can jointly exploit the polar-domain sparsity and the bilinear pattern, it is promising to achieve accurate channel estimation. 
\item 
Finally, we analyze the computational complexity of the BPD based algorithm and compare it with existing algorithms. Extensive simulation results are provided to demonstrate the superior performance of the proposed scheme. It is illustrated that the BPD based channel estimation technique outperforms existing schemes. More importantly, it is verified that the BPD based algorithm is capable of realizing accurate channel estimation in all far-field/near-field/narrowband/wideband conditions\footnote{Simulation codes of this paper are available at
http://oa.ee.tsinghua.edu.cn/dailinglong/publications/publications.html.}. 
\end{itemize}

  \emph{Organization}: The remainder of this paper is organized as follows. Section \ref{sec:2} presents the system model, including the channel model and problem formulation. In Section \ref{sec:3},  the bilinear pattern is proved, and the BPD based algorithm is provided. 
  Simulations are carried out in Section \ref{sec:4}, and finally conclusions are drawn in Section \ref{sec:5}.

\section{System Model}\label{sec:2}

\subsection{Channel Model}
\begin{figure}
  \centering
  \includegraphics[width=3.5in]{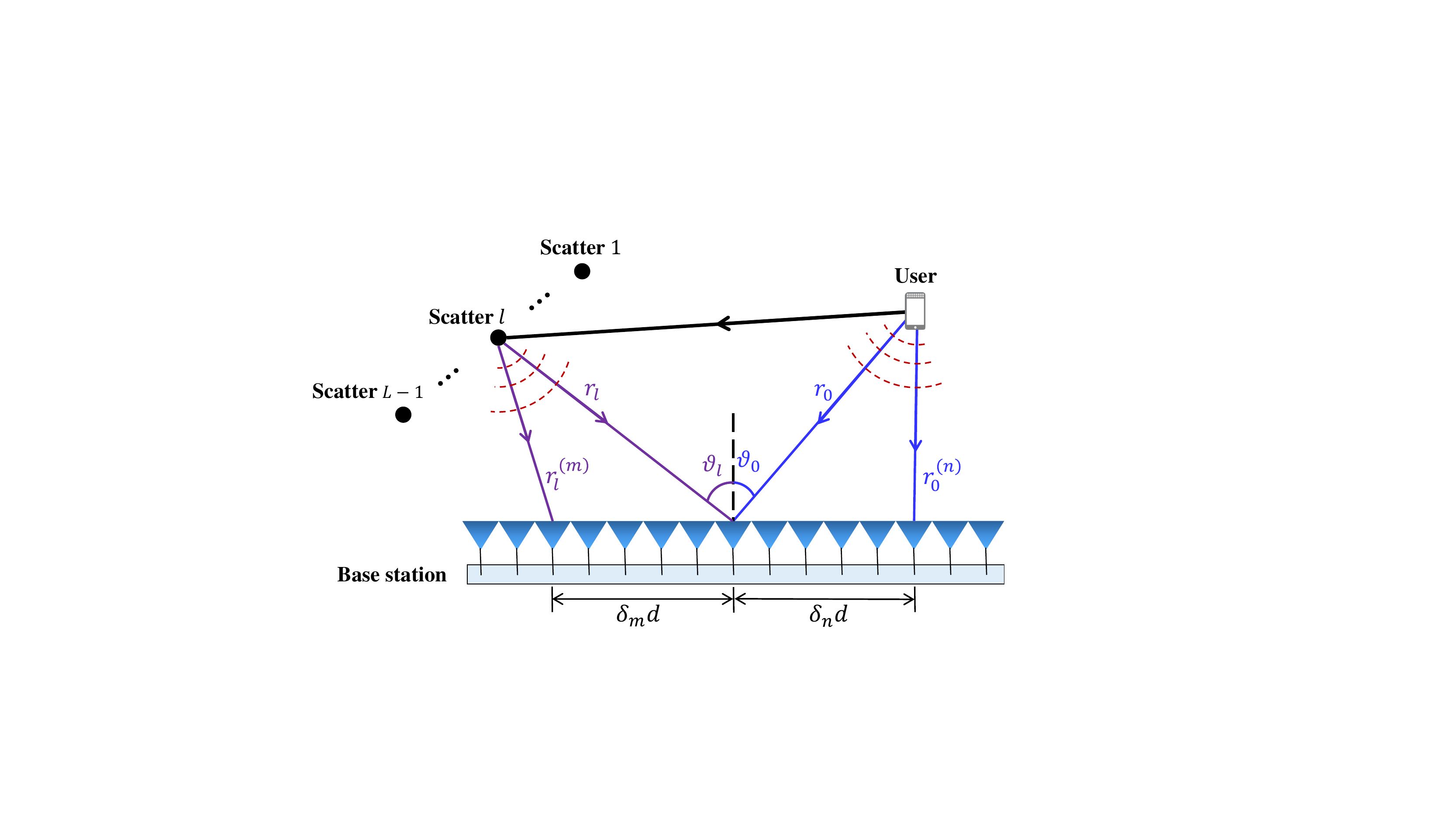}
  \vspace*{-1em}
  \caption{ The near-field channel model for an arbitrary user. 
  }
  \label{fig:channel}
  \vspace*{-1em}
\end{figure}

  In this paper, we investigate the uplink channel estimation for XL-MIMO system with orthogonal frequency division multiplexing (OFDM), where one base station (BS) uses an $N$-antenna uniform linear array to  serve $K$ single-antenna users. We adopt the Saleh-Valenzuela multipath channel to model the near-field channel~\cite{NearLoS_Zhou2015, NearCE_Cui2022}. 
As presented in Fig. \ref{fig:channel}, the frequency-domain channel $\mathbf{h}_{m} \in \mathbb{C}^{N \times 1}$ at the $m$-th subcarrier ($m \in \{0, 1, \cdots, M - 1\}$) from a certain user is denoted as 
\begin{align}\label{eq:Channel}
  \mathbf{h}_{m} = \sqrt{\frac{N}{L}}\sum_{l = 0}^{L - 1} g_{l,m}e^{-j\frac{2\pi}{\lambda_m} r_l }\mathbf{a}(\vartheta_l, r_l, f_m). 
\end{align}
Here, $g_{l,m}$ denotes the complex path gain at $m$-th subcarrier of the $l$-th path, and $L$ represents the number of channel paths. Besides, $f_m = f_c + \frac{2m - M}{2M}B$ is the $m$-th subcarrier frequency with $B$ and $f_c$ corresponding to the bandwidth and the carrier frequency, and $\lambda_m = \frac{c}{f_m}$ is the wavelength at $f_m$ with $c$ being the light speed. Moreover, as shown in Fig. \ref{fig:channel}, $\vartheta_l$ represents the AOA of the $l$-th path, $r_l$ is the distance from the last-hop scatter to the BS array's center, and $\mathbf{a}(\vartheta_l, r_l, f_m)$ is the array response vector of $(\vartheta_l, r_l)$ describing the spherical wavefront between the last-hop scatter and BS. 
According to the geometrical structure of BS array, $\mathbf{a}(\vartheta_l, r_l, f_m)$ is modeled as  
\begin{align} \label{eq:ArrayResponseVector}
\mathbf{a}(\vartheta_l, r_l, f_m) = \frac{1}{\sqrt{N}}[e^{j\phi_{l,m}^{(0)}}, e^{j\phi_{l, m}^{(1)}}, \cdots, e^{j\phi_{l, m}^{(N-1)}}]^T,
\end{align} 
where $ \phi_{l, m}^{(n)} = -\frac{2\pi}{\lambda_m} (r_l^{(n)} - r_l) $ and $r_l^{(n)}$ is the distance from the last-hop scatter to the $n$-th BS antenna. As indicated in~\cite{NearCE_Cui2022}, $r_l^{(n)}$ can be written as $r_l^{(n)} = \sqrt{r_l^2 + \delta_n^2d^2 - 2 r_l \delta_nd \sin\vartheta_l}$, where $\delta_n = n - \frac{N  - 1}{2}$ with $n \in \{0,1,\cdots, N-1\}$. In addition, $d = \frac{\lambda_c}{2}$ is the antena spacing and $\lambda_c = \frac{c}{f_c}$ is the carrier wavelengh. 

From (\ref{eq:ArrayResponseVector}), two crucial aspects are revealed: the near-field propagation and the beam split effect. First, the far-field scenario is always adopted in conventional massive MIMO to simplify $\phi_{l,m}^{(n)}$ as $\frac{2\pi}{\lambda_m}\delta_n d \sin \vartheta_l$, which is independent of distance $r_l$. On the contrary, in XL-MIMO systems, the Fraunhofer distance $\frac{2D^2}{\lambda_c}$ is comparable to $r_l$, making it highly possible for signals to propagate in near-field environments~\cite{fresnel_Selvan2017}. Here, $D = Nd$ is the array aperture. For example, we consider a 0.3-meter-array working at 100 GHz, the Fraunhofer distance of which reaches 60 meters. In this case, the spherical wavefront model has to be investigated and the impact of $r_l$ cannot be ignored~\cite{NearMag_Cui2022}. 
Second, in narrowband communications when $f_m \approx f_c$, the array response vector is nearly frequency-independent, leading to the common sparse structure. This structure allows jointly estimating channels at different subcarriers by using MMV-specific CS algorithms, such as SOMP and the group-sparse Bayesian approach~\cite{SWOMP_Robert18,CE_SOMP_Gao16,CS_Eldar2015,NearCE_Cui2022}. 
However, in wideband communications when $f_m \neq f_c$, the array response vector considerably varies over frequencies. More seriously, the support set of different subcarriers can be quite different from each other in wideband XL-MIMO systems, undermining the common sparse support structure of the MMV model. This phenomenon is defined as the beam split effect~\cite{BSD_Tan21,NFBF_Cui2021}, which severely degrades the performance of conventional channel estimation schemes.

To the best of our knowledge, existing channel estimation works fail to jointly exploit these two underlying characteristics of the wideband XL-MIMO channel, giving rise to poor estimation performance. To fill in this gap, this paper simultaneously considers the near-field propagation and beam split effect (or the near-field beam split effect) to realize high-accuracy channel estimation..

  \subsection{Problem Formulation}
\begin{figure}
  \centering
  \includegraphics[width=4in]{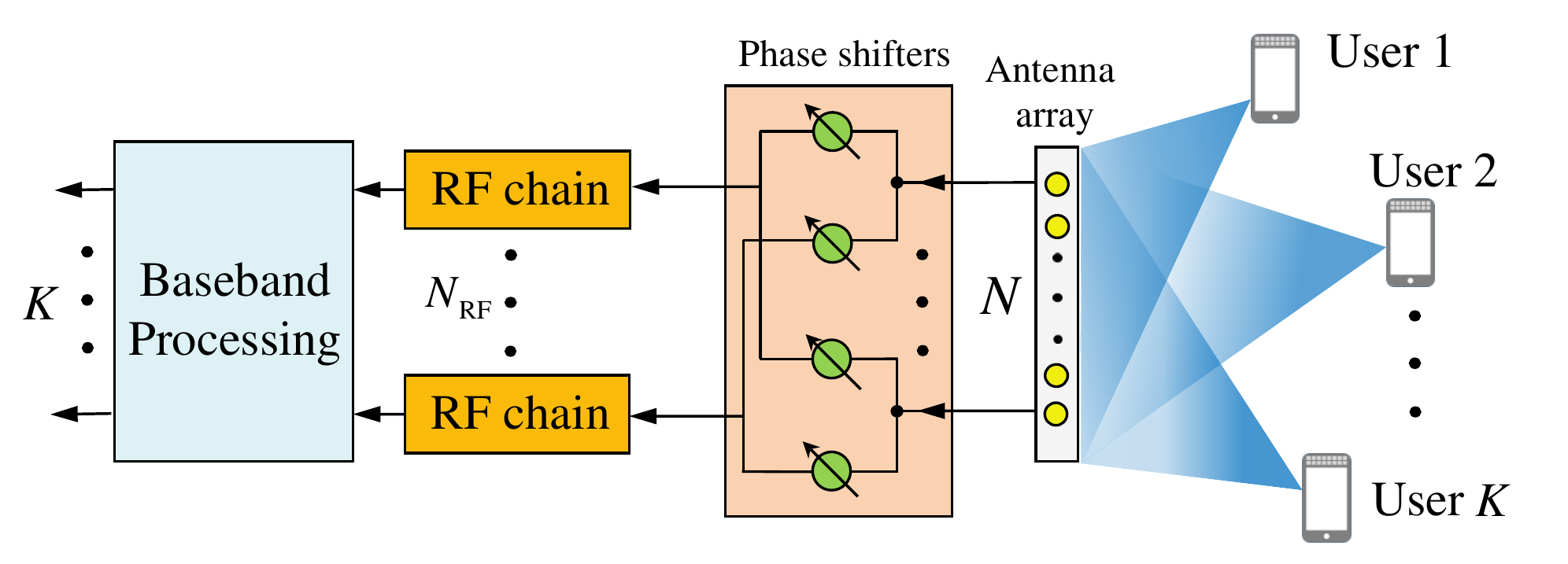}
  \vspace*{-1em}
  \caption{ An illustration of the multi-user uplink transmission system.
  }
  \label{fig:system}
  \vspace*{-1em}
\end{figure}

  As indicated in Fig. \ref{fig:system}, we adopt an uplink time-division-duplex (TDD) channel estimation scenario, where $K$ users harness orthogonal time or frequency resources to 
  transmit orthogonal pilots to BS. Thus we can consider channel estimation for an arbitrary user without loss of generality. 
  We suppose the length of pilot sequence is $P$. Denote $s_{m, p}$ as the pilot signal at the $m$-th subcarrier in the $p$-th time slot. Then, the  received signals $\mathbf{y}_{m, p} \in \mathbb{C}^{N_{\text{RF}} \times 1}$ can be presented as 
  \begin{align}
    \mathbf{y}_{m, p} = \mathbf{A}_p \mathbf{h}_{m} s_{m, p} + \mathbf{A}_p \mathbf{n}_{m,p},
  \end{align}
  where $\mathbf{A}_p \in \mathbb{C}^{N_{\text{RF}} \times N}$ denotes the combining matrix and $\mathbf{n}_{m,p} \in \mathbb{C}^{N \times 1}$ denotes the complex Gaussian noise following the distribution $\mathcal{CN}(0, \sigma^2 \mathbf{I}_N)$. Assume $s_{m, p} = 1$ for $p = 1,2,\cdots, P$. 
  Then, considering the total $P$ time slots for pilot trasmission, we arrive at 
  \begin{align}
    \mathbf{y}_{m} = \mathbf{A} \mathbf{h}_{m} + \mathbf{n}_m,
  \end{align}
  where $\mathbf{y}_m = [\mathbf{y}_{m, 1}^T, \mathbf{y}_{m, 2}^T, \cdots, \mathbf{y}_{m, P}^T]^T \in \mathbb{C}^{PN_{\text{RF}} \times 1}$ and $\mathbf{n}_m = [\mathbf{n}_{m,1}^T \mathbf{A}_1^T, \mathbf{n}_{m,2}^T \mathbf{A}_2^T, \cdots, \mathbf{n}_{m,P}^T \mathbf{A}_P^T]^T \in \mathbb{C}^{PN_{\text{RF}} \times 1}$. Besides, $\mathbf{A} = [\mathbf{A}_{1}^T, \mathbf{A}_{2}^T, \cdots, \mathbf{A}_{P}^T]^T \in \mathbb{C}^{PN_{\text{RF}} \times N}$ is the overall observation matrix, whose elements can be independently generated from the binomial distribution $\frac{1}{\sqrt{N}}\{-1, 1\}$. 
  
  To recover the high-dimensional  near-field channel $\mathbf{h}_{m}$ from the low-dimensional observation $\mathbf{y}_m$, where $PN_{\text{RF}} < N$, we can transform the antenna-domain channel to its angle-distance domain (polar domain) by a polar-domain representation matrix $\mathbf{W} \in \mathbb{C}^{N \times N_aN_d}$~\cite{NearCE_Cui2022}. As discussed in~\cite{NearCE_Cui2022}, $\mathbf{W}$ is composed of $N_d$ submatrices, where 
  \begin{align}
    \mathbf{W} = [\mathbf{W}_1, \mathbf{W}_2, \cdots, \mathbf{W}_{N_d}].
  \end{align} 
  Each submatrix $\mathbf{W}_{n_d} \in \mathbb{C}^{N \times N_a}$ includes $N_a$ array response vectors at the carrier frequency, where 
  \begin{align}\label{eq:Wn}
  \mathbf{W}_{n_d} = [\mathbf{a}(\bar{\vartheta}_1, \bar{r}_{n_d, 1}, f_c), \mathbf{a}(\bar{\vartheta}_2, \bar{r}_{n_d, 2}, f_c), \cdots, \mathbf{a}(\bar{\vartheta}_{n_a}, \bar{r}_{n_d, N_a}, f_c)].
  \end{align}
  As discussed in~\cite{NearCE_Cui2022}, the sampled angles and distances, i.e., ($\bar{\vartheta}_{n_a}, \bar{r}_{n_d, n_a}$) with $n_a \in [1, 2, \cdots, N_a]$ and $n_d \in [1,2, \cdots, N_d]$, should satisfy $\sin\bar{\vartheta}_{n_a} = \frac{2(n_a - 1) - N_a}{N_a}$ and $\bar{r}_{n_d, n_a} = \frac{D^2\cos^2\bar{\vartheta}_{n_a}}{2\beta^2 \lambda_c n_d}$, where $\beta$ is a predefined parameter. Then, the polar-domain channel $\bar{\mathbf{h}}_m$ can be obtained by solving the  underdetermined equation $\mathbf{h}_m = \mathbf{W} \bar{\mathbf{h}}_m$. As investigated in~\cite{NearCE_Cui2022}, matrix $\mathbf{W}$ is capable of extracting the angle-distance information embedded in each array response vector of (\ref{eq:Channel}). In addition, the number of paths $L$ is usually small, especially for mmWave and THz bands (e.g. $L = 5$). As a consequence, the polar-domain channel $\bar{\mathbf{h}}_m$ is sparse and the near-field channel estimation problem is equivalent to a sparse signal recovery problem:
  \begin{align}\label{eq:SparseRecover}
    \mathbf{y}_m = \mathbf{A}\mathbf{W}\bar{\mathbf{h}}_m + \mathbf{n}_m. 
  \end{align}
  Finally, we attempt to jointly estimate channels for all subcarriers. Hence, we rewrite (\ref{eq:SparseRecover}) as 
  \begin{align}\label{eq:MMV}
    \mathbf{Y} = \mathbf{A}\mathbf{W}\bar{\mathbf{H}} + \mathbf{N}, 
  \end{align}
  where $\mathbf{Y} = [\mathbf{y}_1, \mathbf{y}_2, \cdots, \mathbf{y}_M] \in \mathbb{C}^{PN_{\text{RF}} \times M }$, $\bar{\mathbf{H}} = [\bar{\mathbf{h}}_1, \bar{\mathbf{h}}_2, \cdots, \bar{\mathbf{h}}_M]$, and $\mathbf{N} = [\mathbf{n}_1, \mathbf{n}_2, \cdots, \mathbf{n}_M]$. 
  Notice that as the array response vector (\ref{eq:ArrayResponseVector}) is frequency-dependent, the sparse support set of $\bar{\mathbf{h}}_m$ for different subcarriers $f_m$ is quite different from each other. Therefore, existing wideband channel estimation algorithms tailored for MMV problems cannot perfectly match the model (\ref{eq:MMV}). Consequently, an accurate near-field channel estimation scheme for wideband XL-MIMO is essential. 

  \section{Bilinear Pattern Detection Based Channel Estimation} \label{sec:3}
  
  In this section, we first reveal the bilinear pattern of near-field beam split effect. This pattern reveals that with proper preprocessing, the support set of near-field channels varies linearly over frequencies in both the angle and the distance domain. Then, we propose a bilinear pattern detection (BSD) based near-field wideband channel estimation scheme. Finally,  we provide complexity analysis of our scheme. 

  \subsection{Bilinear Pattern of Near-Field Beam Split}
  
  To jointly recover the wideband channel for all subcarriers in (\ref{eq:MMV}), it is crucial to investigate the map between the support index and frequency for an arbitrary path, which is defined as the beam split pattern (BSP). This pattern has been extensively studied under \emph{far-field} environments in the literature~\cite{THzBS_Gao2021,BSD_Tan21,THzCE_Dov2021}, while we are trying to discover the pattern of \emph{near-field beam split}. 

  Specifically, we investigate the $l$-th scatter located at $(\vartheta_l, r_l)$, which corresponds to a series of array response vectors $\mathbf{a}(\vartheta_l, r_l, f_m)$ with different frequencies.  According to the CS theory, the support index $(n_{a, l, m}^{\star}, n_{d, l, m}^{\star})$ of $\mathbf{a}(\vartheta_l, r_l, f_m)$ refers to the index of the strongest element of $\mathbf{W}^H\mathbf{a}(\vartheta_l, r_l, f_m)$, i.e., $(n_{a, l, m}^{\star}, n_{d, l, m}^{\star}) = \arg\max_{n_a, n_d} \|\mathbf{a}^H(\bar{\vartheta}_{n_a}, \bar{r}_{n_d, n_a}, f_c)\mathbf{a}(\vartheta_l, r_l, f_m)\|$.  Unfortunately, as the phase $\phi_{l, m}^{(n)} = -\frac{2\pi}{\lambda_m} (r_l^{(n)} - r_l)$ of the $n$-th element of $\mathbf{a}(\vartheta_l, r_l, f_m)$ is a complicated radical function, it is intractable to obtain the close form of $(n_{a, l, m}^{\star}, n_{d, l, m}^{\star})$. 

  To address this problem, we use the second-order Taylor expansion to approximate $\phi_{l, m}^{(n)}$ as 
  \begin{align}
  \phi_{l, m}^{(n)} \approx \bar{\phi}_{l, m}^{(n)} = \frac{2\pi}{\lambda_m}(\delta_n d \sin\vartheta_l - \delta_n^2 d^2 \frac{\cos^2\vartheta_l}{2r_l}),
  \end{align}
  which is much more accurate than the conventional far-field approximation with first-order Taylor expansion. For expression simplicity, we use parameters $(\theta_l, \alpha_l) = (\sin\vartheta_l, \frac{\cos^2\vartheta_l}{2r_l})$ to replace parameters $(\vartheta_l, r_l)$, so we have $\bar{\phi}_{l, m}^{(n)} = \frac{2\pi}{\lambda_m}(\delta_n d \theta_l - \delta_n^2 d^2 \alpha_l) $. As a result, we can use $\bar{\mathbf{a}}(\theta_l, \alpha_l, f_m) = {\frac{1}{\sqrt{N}}}[e^{j\bar{\phi}_{l,m}^{(0)}}, \cdots, e^{j\bar{\phi}_{l, m}^{(N-1)}}]^T$ to simplify $\mathbf{a}(\vartheta_l, r_l, f_m)$. 
  Similarly, vector $\mathbf{a}(\bar{\vartheta}_{n_a}, \bar{r}_{n_d, n_a}, f_c)$ can be approximated as $\bar{\mathbf{a}}(\bar{\theta}_{n_a}, \bar{\alpha}_{n_d}, f_c)$. Notice that the sampled angles $\bar{\vartheta}_{n_a}$ and distances $\bar{r}_{n_d, n_a}$ are transformed to $\bar{\theta}_{n_a} = \sin\bar{\vartheta}_{n_a} = \frac{2(n_a - 1) - N_a}{N_a}$ and $\alpha_{n_b} = \frac{\cos^2\bar{\vartheta}_{n_a}}{2\bar{r}_{n_d, n_a}} = \frac{\cos^2\bar{\vartheta}_{n_a}}{2 \frac{D^2\cos^2\bar{\vartheta}_{n_a}}{2\beta^2 \lambda_c n_d}} = \frac{\beta^2 \lambda_c n_d}{D^2}$, with $n_a \in \{1, 2, \cdots, N_a\}$ and $n_d \in \{1, 2, \cdots, N_d\}$. 

  Based on the above discussion, the near-field beam split pattern can be acquired by solving the problem $(n_{a, l, m}^{\star}, n_{d, l, m}^{\star}) = \arg\max_{n_a, n_d} \|\bar{\mathbf{a}}^H(\bar{\theta}_{n_a}, \bar{\alpha}_{n_d}, f_c)\bar{\mathbf{a}}(\theta_l, \alpha_l, f_m)\|$. The following \textbf{Lemma 1} provides a more concise form of $\|\bar{\mathbf{a}}^H(\bar{\theta}_{n_a}, \bar{\alpha}_{n_d}, f_c)\bar{\mathbf{a}}(\theta_l, \alpha_l, f_m)\|$.  
\lemma{
  We consider two frequencies $f_1$ and $f_2$, and two arbitrary locations $(\vartheta_1, r_1)$ and $(\vartheta_2, r_2)$ with $(\theta_1, \alpha_1) = (\sin\vartheta_1, \frac{\cos^2\vartheta_1}{2r_1})$ and $(\theta_2, \alpha_2) = (\sin\vartheta_2, \frac{\cos^2\vartheta_2}{2r_2})$. When the number of antennas $N \rightarrow +\infty$, then $\|\bar{\mathbf{a}}^H(\theta_1, \alpha_1, f_1)\bar{\mathbf{a}}(\theta_2, \alpha_2, f_2)\|$ can be represented by 
\begin{align}\label{eq:Lemma1}
  \|\bar{\mathbf{a}}^H(\theta_1, \alpha_1, f_1)\bar{\mathbf{a}}(\theta_2, \alpha_2, f_2)\| \overset{N \rightarrow +\infty}{=}  \left| 
        \int_{-\frac{1}{2}}^{\frac{1}{2}} e^{j2\pi x\gamma - 
      j2\pi x^2\zeta } {\rm d} x
      \right| = \Xi(\gamma, \zeta), 
\end{align}
where $\gamma = \frac{D}{\lambda_1}(\theta_1 - \frac{f_2}{f_1}\theta_2)$, 
$\zeta = \frac{D^2}{\lambda_1}(\alpha_1 - \frac{f_2}{f_1}\alpha_2)$, and $\lambda_1 = \frac{c}{f_1}$.
}

  \begin{figure}
  \centering
  \includegraphics[width=5in]{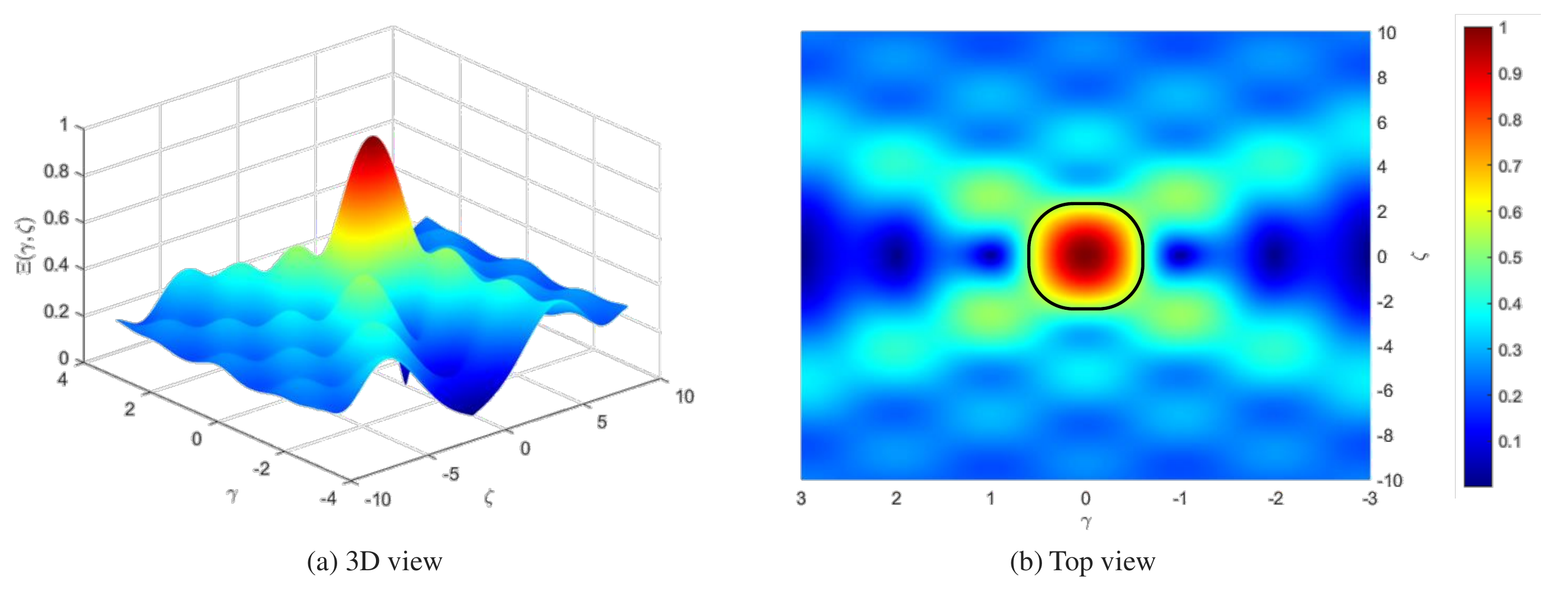}
  \vspace*{-1em}
  \caption{ The numerical results of function $\Xi(\gamma, \zeta)$.
  }
  \label{fig:Xi}
  \vspace*{-1em}
\end{figure}
\emph{Proof}:   Accoding to the definition of $\bar{\mathbf{a}}(\theta_1, \alpha_1, f_1)$ and $\bar{\mathbf{a}}(\theta_2, \alpha_2, f_2)$, we have 
    \begin{align}
      \|\bar{\mathbf{a}}^H(\theta_1, \alpha_1, f_1)\bar{\mathbf{a}}(\theta_2, \alpha_2, f_2)\| &= \left|\frac{1}{N}\sum_{n = -(N-1)/2}^{(N-1)/2}e^{j\frac{2\pi nd}{\lambda_1}(\theta_1 - \frac{f_2}{f_1}\theta_2) - 
      j\frac{2\pi n^2d^2}{\lambda_1}(\alpha_1 - \frac{f_2}{f_1}\alpha_2)} \right| \notag \\
      & \overset{(a)}{=} \left|\sum_{x = -\frac{1}{2} + \frac{1}{2N}}^{\frac{1}{2} - \frac{1}{2N}}e^{j\frac{2\pi Dx}{\lambda_1}(\theta_1 - \frac{f_2}{f_1}\theta_2) - 
      j\frac{2\pi D^2x^2}{\lambda_1}(\alpha_1 - \frac{f_2}{f_1}\alpha_2)} \frac{1}{N}\right|, \label{eq:Summation}
    \end{align}
    where (a) is achieved by letting $x = \frac{n}{N}$ and $D = Nd$. In addition, since XL-MIMO has a very large number of antennas $N$, the summation (\ref{eq:Summation}) can be rewriten as the following Riemann integral 
      \begin{align}\label{eq:Integral}
      \|\bar{\mathbf{a}}^H(\theta_1, \alpha_1, f_1)\bar{\mathbf{a}}(\theta_2, \alpha_2, f_2)\| \overset{N  \rightarrow +\infty}{=} \left| 
        \int_{-\frac{1}{2}}^{\frac{1}{2}} e^{j\frac{2\pi Dx}{\lambda_1}(\theta_1 - \frac{f_2}{f_1}\theta_2) - 
      j\frac{2\pi D^2x^2}{\lambda_1}(\alpha_1 - \frac{f_2}{f_1}\alpha_2)} \text{d}x
      \right|.
    \end{align}
    Eventually, plug $\gamma = \frac{D}{\lambda_1}(\theta_1 - \frac{f_2}{f_1}\theta_2)$ and $\zeta = \frac{D^2}{\lambda_c}(\alpha_1 - \frac{f_2}{f_1}\alpha_2)$ into (\ref{eq:Integral}), the right-hand side in (\ref{eq:Lemma1}) can be acquired, and the proof is completed. 
$\hfill\blacksquare$

\textbf{Lemma 1} indicates that for XL-MIMO communications with a very large number of antennas $N$, finding the support index $(n_{a, l, m}^{\star}, n_{d, l, m}^{\star})$ is equivalent to solving the following  problem:
\begin{align}\label{eq:Optimization1}
(n_{a, l, m}^{\star}, n_{d, l, m}^{\star}) = \arg\max_{n_a, n_d} 
\Xi\left(\frac{D}{\lambda_c}(\bar{\theta}_{n_a} - \frac{f_m}{f_c}\theta_l), \frac{D^2}{\lambda_c}(\bar{\alpha}_{n_d} - \frac{f_m}{f_c}\alpha_l)\right).
\end{align}
To address problem (\ref{eq:Optimization1}), it is of great importance to grasp the property of function $\Xi(\gamma, \zeta)$. First of all, it is easy to prove that $\Xi(\gamma, \zeta)$ is an even function with respect to (w.r.t) $\gamma$ and $\zeta$, i.e., 
\begin{align}
\Xi(\gamma, \zeta) = \Xi(-\gamma, \zeta), \quad 
\Xi(\gamma, \zeta) = \Xi(\gamma, -\zeta).
\end{align}
Therefore, we can rewrite $\Xi(\gamma, \zeta)$ as $\Xi(\gamma, \zeta) = \Xi(|\gamma|, |\zeta|)$, or we only need to discuss the domain $\gamma > 0$ and $\zeta > 0$.

In addition, $\Xi(\gamma, \zeta)$ is a non-parameter function, we can use numerical integration to acquire its numerical result, which is plotted in Fig. \ref{fig:Xi}. This numerical result indicates that within the main lobe of $\Xi(\gamma, \zeta)$ (area surrounded by the black line in Fig. \ref{fig:Xi} (b)), the value of $\Xi(\gamma, \zeta)$ declines with the increase of $|\gamma|$ and $|\zeta|$. That is to say, if $|\gamma_1| \le |\gamma_2|$ and $|\zeta_1| \le |\zeta_2|$, then we have $\Xi(\gamma_1, \zeta_1) \ge \Xi(\gamma_2, \zeta_2)$. Making use of these properties, \textbf{Lemma 2} gives out the close form solution of (\ref{eq:Optimization1}). 

\lemma{
 If $\frac{D}{\lambda_c}(\theta_{n_{a, l, m}^{\star}} - \frac{f_m}{f_c}\theta_l)$ and $\frac{D^2}{\lambda_c}(\alpha_{n_{a, l, m}^{\star}} - \frac{f_m}{f_c}\alpha_l)$ are within the main lobe of $\Xi(\gamma, \zeta)$, then the optimal solution of problem (\ref{eq:Optimization1}) is 
 \begin{align}
  n_{a, l, m}^{\star} = \arg\min_{{n_a}} |\bar{\theta}_{n_a} - \frac{f_m}{f_c}\theta_l|, \label{eq:s1}\\
  n_{d, l, m}^{\star} = \arg\min_{{n_d}} |\bar{\alpha}_{n_d} - \frac{f_m}{f_c}\alpha_l|. \label{eq:s2}
\end{align}
}
\emph{Proof}: In the main lobe, since $\Xi(|\gamma|, |\zeta|)$ is a decreasing function w.r.t $|\gamma|$ and $|\zeta|$, and $\Xi(\gamma, \zeta) = \Xi(|\gamma|, |\zeta|)$, hence maximizing $\Xi\left(\frac{D}{\lambda_c}(\bar{\theta}_{n_a} - \frac{f_m}{f_c}\theta_l), \frac{D^2}{\lambda_c}(\bar{\alpha}_{n_d} - \frac{f_m}{f_c}\alpha_l)\right)$ is equivalent to minimizing $|\frac{D}{\lambda_c}(\bar{\theta}_{n_a} - \frac{f_m}{f_c}\theta_l)|$ and $|\frac{D^2}{\lambda_c}(\bar{\alpha}_{n_d} - \frac{f_m}{f_c}\alpha_l)|$ separately, which gives rise to the solutions (\ref{eq:s1}) and (\ref{eq:s2}). 
$\hfill\blacksquare$

\begin{figure}
  \centering
  \includegraphics[width=5.5in]{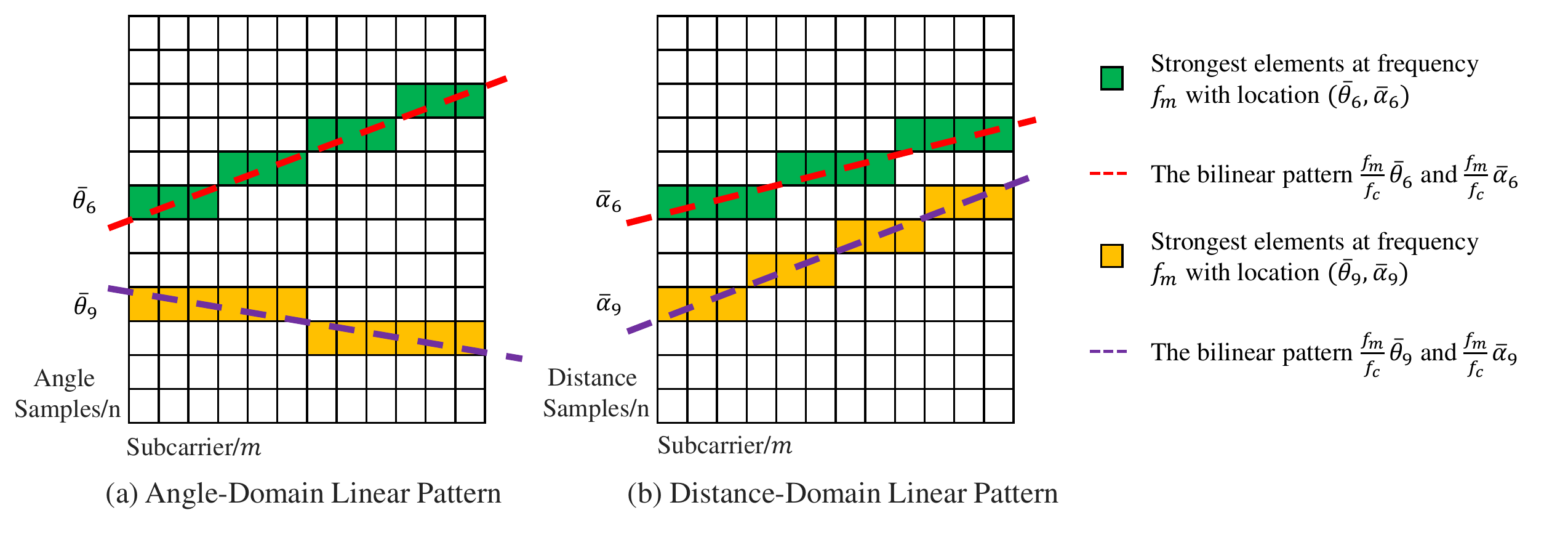}
  \vspace*{-1em}
  \caption{ The bilinear pattern of the near-field beam split effect. 
  }
  \label{fig:bilinear}
  \vspace*{-1em}
\end{figure}

\textbf{Lemma 2} explicitly displays the bilinear pattern over frequencies of the near-field beam split effect. As illustrated by the dotted lines in Fig. \ref{fig:bilinear}, both  $\frac{f_m}{f_c}\alpha$ and $\frac{f_m}{f_c}\theta$ grow linearly to frequency $f_m$. The support indices $n_{a, l, m}^{\star}$  and $n_{d, l, m}^{\star}$ correspond to the sampled $\theta_{n_{a}}$ and $\alpha_{n_d}$ that are nearest to $\frac{f_m}{f_c}\theta$ and $\frac{f_m}{f_c}\alpha$. Thus, this double linear growth pattern is defined as the bilinear pattern of near-field beam split. 
More importantly, the bilinear pattern can be utilized to capture the support set of the $l$-th path for all frequencies, so as to accurately estimate the location $(\vartheta_l, r_l)$.  Its superiority comes from two aspects. At first, \textbf{Lemma 2} implies that the indices $n_{a, l, m}^{\star}$  and $n_{d, l, m}^{\star}$ can gather the largest power of the $l$-th path from the entire bandwidth. 
Therefore, compared to conventional common support set assumption in MMV model, the proposed bilinear pattern enjoys higher recovery accuracy for wideband channels. 
In addition, different from existing works that only explain the beam split pattern (\ref{eq:s1}) in the angle domain~\cite{THzBS_Gao2021,BSD_Tan21,THzCE_Dov2021}, the bilinear pattern reveals the linear-changing structure in the joint angle-distance domain. Thus, the proposed pattern is capable of estimating both far-field and near-field channels.

  \subsection{Bilinear Pattern Detection Based Channel Estimation}

In this subsection, based on the bilinear pattern discussed before, a bilinear pattern detection based channel estimation scheme is proposed to solve problem $(\ref{eq:MMV})$. {\color{black} This algorithm is inspired by the classical CS-based polar-domain SOMP method \cite{NearCE_Cui2022}. As illustrated in \cite{NearCE_Cui2022}, the polar-domain SOMP method assumes that the polar-domain channel $\bar{\mb{h}}_m$ is sparse and the sparse support sets of $\bar{\mb{h}}_m, \forall m$ are the same. Based on this assumption, the polar-domain SOMP method computes the total power of each row of $\bar{\mb{H}}$ so as to accumulate the information from the entire bandwidth. In this way, it can recover the location $(\vartheta_l, r_l)$. However, we have proved previously that the actual sparse support sets for different frequencies are not the same in wideband conditions. Indeed, they should satisfy the bilinear pattern presented in (\ref{eq:s1})-(\ref{eq:s2}). Therefore, instead of computing the total power of each row of $\bar{\mb{H}}$, our BPD based algorithm takes advantage of the bilinear pattern to collect powers from all frequencies, so as to improve the accuracy of the estimation for $(\vartheta_l, r_l)$.
}

\begin{algorithm}
\footnotesize
\caption{Bilinear pattern detection based channel estimation algorithm}
\label{alg1}
\begin{algorithmic}[1]
    \REQUIRE Received signal matrix $\mathbf{Y}$; Observation matrix $\mathbf{A}$; Polar-domain representation matrix $\mathbf{W}$; Number of paths to be detected $\hat{L}$;
    \ENSURE Estimated wideband channel $\hat{\mathbf{H}}$;\\
    \textbf{Pre-whitening stage} \\
    \STATE $\mathbf{C} = {\rm{diag}}\{\mathbf{A}_1\mathbf{A}_1^H, \mathbf{A}_2\mathbf{A}_2^H, \cdots, \mathbf{A}_P\mathbf{A}_P^H\}$; 
    \STATE $\mathbf{C} =  \sigma^2 \mb{S}\bm{\Sigma}\mb{S}^H$, $\mb{D} = \mb{S}\bm{\Sigma}^{\frac{1}{2}}$;
    \STATE Use matrix $\mathbf{D}$ to pre-white the received signal as $\bar{\mathbf{Y}} = \mathbf{D}^{-1}\mathbf{Y}  = \bm{\Psi}\bar{\mathbf{H}} + \bar{\mathbf{N}} = [\bar{\mb{y}}_1, \cdots, \bar{\mb{y}}_M]$;\\
    \textbf{Path detection stage} \\
    \STATE Initialize the residue matrix as $\mathbf{R} = [\mathbf{r}_1, \mathbf{r}_2, \cdots, \mathbf{r}_M] = \bar{\mathbf{Y}}$ and the support set as $\Upsilon = \{\emptyset\}$; 
    \STATE Angle-domain linear pattern: $\Gamma({n_a}, m) = \arg$\!$\min_n |\bar{\theta}_n - \frac{f_m}{f_c}\bar{\theta}_{n_a}|$ for $n_a \in \{ 1,\cdots, N_a\}$ and $m\in \{1,\cdots, M\}$;
    \STATE Distance-domain linear pattern: $\Lambda({n_d}, m) = \arg$\!$\min_n |\bar{\alpha}_n - \frac{f_m}{f_c}\bar{\alpha}_{n_d}|$ for $n_d \in \{ 1,\cdots, N_d\}$ and $m\in \{1,\cdots, M\}$;
    \FOR{$l \in \{1,2,\cdots, \hat{L}\}$}
        \STATE $\mathbf{U} = \bm{\Psi}^H\mathbf{R} = [\mathbf{u}_1, \mathbf{u}_2, \cdots, \mathbf{u}_M]$;
        \STATE $(n_{a, l}^{\star}, n_{d,l}^{\star}) = \arg\max_{(n_a, n_d)} \sum_{m = 1}^M \|u_m\left( (\Lambda(n_d, m) - 1) N_a + \Gamma({n_a}, m) \right) \|^2$;
        \STATE $\Upsilon = \Upsilon \cup \{(n_{a, l}^{\star}, n_{d, l}^{\star})\}$;
        \FOR {$m \in \{1,\cdots, M\}$}
              \STATE $\Upsilon_m = \{ \left(\Lambda(n_d, m) - 1 \right)N_a + \Gamma({n_a}, m) | (n_a, n_d) \in \Upsilon\}$;
              \STATE $\hat{\bar{\mb{h}}}_m = \mb{0}_{N_aN_d \times 1}$,  $\hat{\bar{\mb{h}}}_m(\Upsilon_m) = \bm{\Psi}^{\dagger}(:, \Upsilon_m)\bar{\mb{y}}_m$;
        \ENDFOR
        \STATE Update the residue $\mb{R} = \bar{\mb{Y}} - \bm{\Psi}[\hat{\bar{\mb{h}}}_1, \hat{\bar{\mb{h}}}_2, \cdots, \hat{\bar{\mb{h}}}_M]$;
    \ENDFOR
            \STATE $\hat{\mb{H}} = \mb{W}[\hat{\bar{\mb{h}}}_1, \hat{\bar{\mb{h}}}_2, \cdots, \hat{\bar{\mb{h}}}_M]$.
\end{algorithmic}
\end{algorithm}

Specifically, the pseudo-code of the proposed BPD algorithm is provided in \textbf{Algorithm \ref{alg1}}. The overall algorithm is composed of two stages. In the first pre-whitening stage, as the additional noise $\mb{n}_m$ is colored noise with a covariance matrix of $\mb{C} = \sigma^2\text{diag}\{\mb{A}_1\mb{A}_1^H, \mb{A}_2\mb{A}_2, \cdots, \mb{A}_P\mb{A}_P^H\}$, we use a matrix $\mb{D} \in \mathbb{C}^{PN_{\text{RF}} \times PN_{\text{RF}}}$ to pre-white the received signals. Here, $\mb{D}$ is acquired by calculating the  eigenvalue decomposition $\mb{C} = \sigma^2 \mb{S}\bm{\Sigma}\mb{S}^H$, where $\mb{D} = \mb{S}\bm{\Sigma}^{\frac{1}{2}}$.  Left multiplying the received signals $\mb{Y}$ by $\mb{D}^{-1}$, we have 
\begin{align}
  \bar{\mb{Y}} = \mb{D}^{-1}\mb{Y} = \mb{D}^{-1}\mb{A}\mb{W}\bar{\mb{H}} + \mb{D}^{-1}\mb{N} = \bm{\Psi} \bar{\mb{H}} + \bar{\mb{N}}, 
\end{align}
where $\bm{\Psi} = \mb{D}^{-1}\mb{A}\mb{W}$ and $\bar{\mb{N}} = \mb{D}^{-1}\mb{N}$. Notice that here, the covariance of each column of $\bar{\mb{N}}$ becomes $\mb{D}^{-1}\mb{C}\mb{D}^{-H} = \sigma^2 \bm{\Sigma}^{-\frac{1}{2}}\mb{S}^{-1}\mb{S}\bm{\Sigma}\mb{S}^H\mb{S}^{-H}\bm{\Sigma}^{-\frac{1}{2}} = \sigma^2 \mb{I}_{PN_{\text{RF}}}$, which is white noise.  

In the second stage of path detection, we first initialize the residue matrix $\mb{R} = \mathbb{C}^{PN_{\text{RF}}\times M}$ as $\mb{R} = \bar{\mb{Y}}$. The support set $\Upsilon$ is initialized as $\Upsilon = \{ \emptyset \}$, which will be used to store the detected support indices at carrier frequency $f_c$. 

In steps 5-6, according to (\ref{eq:s1}) and (\ref{eq:s2}), we generate the angle-domain and distance-domain linear patterns as 
\begin{align}
\Gamma(n_a, m) = \arg\min_n |\bar{\theta}_n - \frac{f_m}{f_c}\bar{\theta}_{n_a}|, \\
\Lambda(n_d, m) = \arg\min_n |\bar{\alpha}_n - \frac{f_m}{f_c}\bar{\alpha}_{n_d}|, 
\end{align} 
where $n_a \in \{1,\cdots, N_a\}$, $n_d \in \{1, \cdots, N_d\}$, and $m \in \{1,\cdots, M\}$. Here, all parameters $\bar{\theta}_{n_a}$, $\bar{\theta}_{n}$, $\bar{\alpha}_{n_d}$, and $\bar{\alpha}_{n}$ belong to the sampled locations in $\mb{W}$. $(\Gamma(n_a, m), \Lambda(n_d, m))$ denotes the support index of $(\bar{\theta}_{n_a}, \bar{\alpha}_{n_d})$ at frequency $f_m$. 
To elaborate on steps 5-6, based on the CS theory, our target is to find out a polar-domain sampled location $(\bar{\vartheta}_{n_{a, l}^{\star}}, \bar{r}_{n_{d, l}^{\star}, {n_{a, l}^{\star}}})$  that is nearest to the physical location $(\vartheta_l, r_l)$. 
As indicated in \textbf{Lemma 2}, each sampled location $(\bar{\vartheta}_{n_a}, \bar{r}_{n_d, n_a})$ corresponds to a certain bilinear pattern over frequencies. Hence, we use $\Gamma({n_a}, m) $ and $\Lambda({n_d}, m) $ to represent these patterns.  Then, $\Gamma({n_a}, m)$ and $\Lambda({n_d}, m)$ allow us to accumulate the coherence between $(\bar{\theta}_{n_a}, \bar{\alpha}_{n_d})$ and a physical location $(\vartheta_l, r_l)$ from the entire bandwidth.  
The nearest sample  $(\bar{\vartheta}_{n_{a, l}^{\star}}, \bar{r}_{n_{d, l}^{\star}, {n_{a, l}^{\star}}})$ or $(\bar{\theta}_{n_{a, l}^{\star}}, \bar{\alpha}_{n_{d, l}^{\star}})$ can be selected by capturing the largest coherence.

To be more specific, for the $l$-th path component, the physical location $(\vartheta_l, r_l)$ is estimated relying on the above idea. Inspired by the OMP and SOMP methods, we use the product of $\bm{\Psi}^H$ and $\mb{R}$ to calculate the correlation matrix $\mb{U} = \bm{\Psi}^H\mb{R} = [\mathbf{u}_1, \mathbf{u}_2, \cdots, \mathbf{u}_M] \in \mathbb{C}^{N_aN_d \times M}$ in step 8. After that, in step 9, the bilinear pattern is utilized to capture the power of the correlation matrix for all frequencies. For each sampled $(\bar{\vartheta}_{n_a}, \bar{r}_{n_d, n_a})$ with a support index $(n_a, n_d)$ at carrier frequency $f_c$, its support index becomes $(\Gamma({n_a}, m), \Lambda({n_d}, m))$ at frequency $f_m$. Besides, as the dimension of each column $\mb{u}_m$ of $\mb{U}$ is $N_aN_d \times 1$, the support index $(\Gamma({n_a}, m), \Lambda({n_d}, m))$ corresponds to the $\left( (\Lambda(n_d, m) - 1) N_a + \Gamma({n_a}, m) \right)$-th element of $\mb{u}_m$. As a consequence, the power of the correlation matrix for location $(\bar{\vartheta}_{n_a}, \bar{r}_{n_d,n _a})$ can be written as $\sum_{m = 1}^M \|u_m\left( (\Lambda(n_d, m) - 1) N_a + \Gamma({n_a}, m) \right) \|^2$, and thus the optimal support index of the $l$-th path component is determined as  
\begin{align}
(n_{a, l}^{\star}, n_{d,l}^{\star}) = \arg\max_{(n_a, n_d)} \sum_{m = 1}^M \|u_m\left( (\Lambda(n_d, m) - 1) N_a + \Gamma({n_a}, m) \right) \|^2. 
\end{align}

After detecting $(n_{a, l}^{\star}, n_{d,l}^{\star})$, we add it to the support set $\Upsilon$ in step 10.  
Basically, $\Upsilon$ represents the detected sparse channel supports at the carrier frequency. Then in step 12, $\Upsilon$ is transformed to the sparse channel support set $\Upsilon_m$ at frequency $f_m$. We define each element in $\Upsilon_m$ as the column index of a sparse support index  of $\bar{\mb{h}}_m$, which can be written as 
\begin{align}
  \Upsilon_m = \{ \left(\Lambda(n_d, m) - 1 \right)N_a + \Gamma({n_a}, m) | (n_a, n_d) \in \Upsilon\}. 
\end{align}

Next, in step 13, we denote $\hat{\bar{\mb{h}}}_m$ as the estimated polar-domain channel at frequency $f_m$. Its non-zero elements can be calculated by using $\Upsilon_m$ through the LS algorithm as 
\begin{align}
  \hat{\bar{\mb{h}}}_m = \mb{0}_{N_a N_d \times 1}, \quad \hat{\bar{\mb{h}}}_m(\Upsilon_m) = \bm{\Psi}^{\dagger}(:, \Upsilon_m)\bar{\mb{y}}_m. 
\end{align}
Here, matrix $\bm{\Psi}(:, \Upsilon_m)$ is composed of the column vectors in  $\bm{\Psi}$ indexed by the set $\Upsilon_m$. 
After that, we remove the impact of all detected $l$ paths in the current iteration from the received signals to update the residue matrix as 
\begin{align}
  \mb{R} = \bar{\mb{Y}} - \bm{\Psi}[\hat{\bar{\mb{h}}}_1, \hat{\bar{\mb{h}}}_2, \cdots, \hat{\bar{\mb{h}}}_M].
\end{align}
Steps 8-15 discussed above are repeated $\hat{L}$ times until all path components are detected. Eventually, in step 17, the antenna-domain wideband channel $\hat{\mb{H}}$ is recovered based on these sparse polar-domain channels as 
\begin{align}
  \hat{\mb{H}} = \mb{W}[\hat{\bar{\mb{h}}}_1, \hat{\bar{\mb{h}}}_2, \cdots, \hat{\bar{\mb{h}}}_M].
\end{align}

It is worth mentioning that although the proposed algorithm is inspired by the conventional CS-based methods, it is able to exploit both the near-field and beam split effects embedded in the channel model while existing methods cannot. 
To be specific, existing far-field wideband channel estimation algorithms rely on the angle-domain sparsity and only exploit the angle-domain linear pattern $\Gamma(n_a, m)$, thus these methods will undergo a serious performance loss in the near-field environments. Moreover, existing near-field narrowband channel estimation algorithms certainly take advantage of the polar-domain sparsity, but they assume the channel sparse support sets are the same for all frequencies, which will result in poor estimation accuracy when the bandwidth is large.     On the contrary, the proposed algorithm takes good advantage of the frequency-dependent sparse structure of near-field wideband channels. Leveraging such a bilinear pattern, our scheme is expected to achieve higher estimation accuracy in wideband XL-MIMO systems, which will be demonstrated in the simulation section\footnote{\color{black} Notice that the proposed scheme can achieve higher accuracy on the estimation of the location parameter $(\vartheta_l, r_l)$ at all frequencies than  existing channel estimation schemes. Thus, our scheme can also be used to improve the time of arrival (ToA) estimation accuracy.}. 


\subsection{Complexity Analysis}

\begin{table}[!t]\color{black}
\footnotesize
\caption{\color{black} Computational Complexity}
\label{tab1}
\tabcolsep 30pt 
\begin{tabular*}{\textwidth}{cc}
\toprule
Algorithm                           & Computiational Complexity $\mathcal{O}(\cdot)$ \\ \hline
Angle-domain SOMP \cite{SWOMP_Robert18}          & $\mathcal{O}(\hat{L}N_a P N_{\text{RF}}M) + \mathcal{O}(\hat{L}^3MPN_{RF} + \hat{L}^4) + \mathcal{O}(\hat{L}^2PN_{\text{RF}}M)$      \\ 
BSPD algorithm \cite{BSD_Tan21}             & $\mathcal{O}(\hat{L}N_a P N_{\text{RF}}M) + \mathcal{O}(\hat{L}^3MPN_{RF} + \hat{L}^4) + \mathcal{O}(\hat{L}^2PN_{\text{RF}}M)$         \\ 
Polar-domain SOMP \cite{NearCE_Cui2022}           & $\mathcal{O}(\hat{L}N_aN_d P N_{\text{RF}}M) + \mathcal{O}(\hat{L}^3MPN_{RF} + \hat{L}^4) + \mathcal{O}(\hat{L}^2PN_{\text{RF}}M)$       \\ 
Proposed BPD algorithm      & $\mathcal{O}(\hat{L}N_aN_d P N_{\text{RF}}M) + \mathcal{O}(\hat{L}^3MPN_{RF} + \hat{L}^4) + \mathcal{O}(\hat{L}^2PN_{\text{RF}}M)$     \\ 
\bottomrule
\end{tabular*}
\end{table}

{\color{black}
In this subsection, the computational complexity of the BPD based channel estimation algorithm is analyzed, where we mainly count the number of complex multiplications. As indicated in \textbf{Algorithm 1}, the complexity mainly comes from the iteration procedure of steps 8, 9, 13, and 15. 

In step 8, the product of matrices $\bm{\Psi}^H \in \mathbb{C}^{N_aN_d \times P N_{\text{RF}}}$ and  $\mb{R} \in \mathbb{C}^{P N_{\text{RF}} \times M}$ has a complexity in the order of $\mathcal{O}(N_aN_d P N_{\text{RF}}M)$. Notice that the number of sampled angles $N_a$ and distances $N_d$ is usually proportional to the number of antennas $N$ \cite{NearCE_Cui2022}.

In step 9, we need to calculate the power $\sum_{m = 1}^M \|u_m\left( (\Lambda(n_d, m) - 1) N_a + \Gamma({n_a}, m) \right) \|^2$ for $N_a N_d$ times, whose complexity is in the order of $\mathcal{O}(N_aN_dM)$. 

The complexity of step 13 is dominated by the calculation of matrix ${\bm{\Psi}}^{\dagger}(:, \Upsilon_m) \in \mathbb{C}^{l \times PN_{\text{RF}}}$, which can be regarded as ${\bm{\Psi}}^{\dagger}(:, \Upsilon_m) = ({\bm{\Psi}}^H(:, \Upsilon_m){\bm{\Psi}}(:, \Upsilon_m))^{-1}{\bm{\Psi}}^H(:, \Upsilon_m)$. 
Hence, the complexity of calculating ${\bm{\Psi}}^{\dagger}(:, \Upsilon_m)$ consists of three computation steps: the product of ${\bm{\Psi}}^H(:, \Upsilon_m)$ and ${\bm{\Psi}}(:, \Upsilon_m)$, the inverse of ${\bm{\Psi}}^H(:, \Upsilon_m){\bm{\Psi}}(:, \Upsilon_m)$, and the product of $({\bm{\Psi}}^H(:, \Upsilon_m){\bm{\Psi}}(:, \Upsilon_m))^{-1}$ and ${\bm{\Psi}}^H(:, \Upsilon_m)$.  Therefore, step 13 has a complexity in the order of $\mathcal{O}(l^2P N_{RF} + l^3 + l^2PN_{RF}) = \mathcal{O}(l^2PN_{RF} + l^3)$. 
As step 13 is carried out $M$ times, its overall complexity is in the order of $\mathcal{O}(M(l^2PN_{RF} + l^3))$. 

Next, in step 15, each column vector $\hat{\bar{\mb{h}}}_m$ only has $l$ non-zero elements. As a result, the left multiplying $[\hat{\bar{\mb{h}}}_1, \hat{\bar{\mb{h}}}_2, \cdots, \hat{\bar{\mb{h}}}_M]$  by matrix $\bm{\Psi}$ has a complexity of $\mathcal{O}(PN_{\text{RF}}lM)$. 

Eventually, steps 8, 9, 13, 15 are executed $\hat{L}$ times with $l = 1,2,\cdots, \hat{L}$. Therefore, the total complexity can be summarized as 
\begin{align}\label{eq:comp}
  &\:\mathcal{O}(\hat{L}N_aN_d P N_{\text{RF}}M) + \mathcal{O}(\hat{L}N_aN_dM) + \mathcal{O}(\hat{L}^3MPN_{RF} + \hat{L}^4) + \mathcal{O}(\hat{L}^2PN_{\text{RF}}M) \notag \\
  = &\:\mathcal{O}(\hat{L}N_aN_d P N_{\text{RF}}M) + \mathcal{O}(\hat{L}^3MPN_{RF} + \hat{L}^4) + \mathcal{O}(\hat{L}^2PN_{\text{RF}}M).
\end{align}

	In Table \ref{tab1}, we have provided the computational complexity comparison for different channel estimation algorithms, including the angle-domain SOMP algorithm \cite{SWOMP_Robert18} for far-field narrowband channel estimation, the BSPD algorithm \cite{BSD_Tan21}  for far-field wideband channel estimation, and the polar-domain SOMP algorithm \cite{NearCE_Cui2022}  for near-field narrowband channel estimation. 
	It is clear from Table \ref{tab1} that the proposed BPD algorithm has the same complexity as the polar-domain SOMP scheme in \cite{NearCE_Cui2022}. This is because the main difference between these two algorithms lies in step 9. Specifically, the polar-domain SOMP technique uses the common sparse support set assumption to accumulate  the power while the proposed algorithm exploits the bilinear pattern to accumulate the power, both of which have the same complexity $\mathcal{O}(N_aN_dM)$.  

	Moreover, the complexity of near-field algorithms is higher than that of far-field algorithms, since the far-field methods only need to recover the AoA parameters while the near-field schemes have to detect the AoA and distance parameters simultaneously. 
	To illustrate, the complexity difference comes from the first item in (\ref{eq:comp}), i.e., $\mathcal{O}(\hat{L}N_aN_d P N_{\text{RF}}M)$ for near-field methods and $\mathcal{O}(\hat{L}N_d P N_{\text{RF}}M)$ for far-field methods. 
	As the far-field channel estimation solely cares about the AoA information, the number of sampled distances $N_d$ in \cite{SWOMP_Robert18, BSD_Tan21} can be regarded as 1, giving rise to the complexity of $\mathcal{O}(\hat{L}N_d P N_{\text{RF}}M)$. Despite the lower computational complexity, the far-field algorithms neglect the essential distance information, so they can hardly achieve accurate XL-MIMO channel estimation, which is demonstrated in the next section.

}
  \section{Simulation Results}\label{sec:4}

\begin{table}[!t]
\footnotesize
\caption{Simulation Configurations}
\label{tab2}
\tabcolsep 15pt 
\begin{tabular*}{\textwidth}{cccc}
\toprule
Parameter                           & Value   & Parameter                                & Value \\ \hline
Number of BS antennas $N$           & 256     & Number of angle-domain samples $N_a$     & 256   \\ 
Number of RF chains $N_{\text{RF}}$ & 4       & Number of distance-domain samples $N_d$  & 14    \\ 
Number of users $K$                 & 4       & Parameter $\beta$                        & 0.8   \\ 
Number of subcarriers $M$           & 256     & Number of channel paths $L$              & 6     \\ 
Carrier frequency $f_c$             & 100 GHz & Number of paths to be detected $\hat{L}$ & 12    \\ 
The distribution of path gain $g_{l,m}$    & $\mathcal{CN}(0, 1)$ & The distribution of angle $\vartheta_l$ & $\mathcal{U}\left( -\frac{\pi}{2}, \frac{\pi}{2} \right)$    \\ 
\bottomrule
\end{tabular*}
\end{table}

In this section, we present simulation results to demonstrate the performance of the proposed algorithm.  A wideband XL-MIMO system is considered, and some of the simulation configurations are shown in Table \ref{tab2}. {\color{black} According to these configurations, the number of BS's antennas is $N = 256$ and the carrier wavelength is $\lambda_c = \frac{c}{f_c} = 3$ mm. Therefore, the antenna spacing is $d = \frac{\lambda_c}{2} = 1.5$ mm, and the BS array aperture is $D = (N-1)d \approx N d = \frac{N\lambda_c}{2} = 0.384$ m. Then, it can be derived that the Fraunhofer distance is $\frac{2D^2}{\lambda_c} = 98.3$ m. }
In addition, the distance parameters $r_l$ in (\ref{eq:Channel}) are randomly generated from the uniform distribution $\mathcal{U}(R_{\min}, R_{\max})$, where the smallest distance $R_{\min}$ and the largest distance $R_{\min}$ have different values for different simulations. The signal-to-noise radio is defined as $\text{SNR} = \mathbb{E}\left( \| \mb{H} \|_F^2 / \| \bar{\mb{N}} \|_F^2\right)$. 
The compared benchmark channel estimation algorithms are as follows: 1) the LS algorithm; 2) the far-field narrowband channel estimation schemes, including the angle-domain OMP~\cite{CE_OMP_Lee16} and SOMP~\cite{SWOMP_Robert18} algorithms; 3) the BSPD based far-field wideband channel estimation~\cite{BSD_Tan21}; 4) the near-field narrowband channel estimation schemes, including the polar-domain OMP and SOMP algorithms~\cite{NearCE_Cui2022}.   We use the performance of normalized mean square error (NMSE) to evaluate different algorithms, which is defined as 
$\text{NMSE} = \mathbb{E} \left( \frac{\| \mb{H} - \hat{\mb{H}}\|_F^2 } {\| \mb{H} \|_F^2} \right)$. Besides, 300 Monte Carlo experiments are carried out to plot each figure. 

\begin{figure}
  \centering
  \includegraphics[width=3.5in]{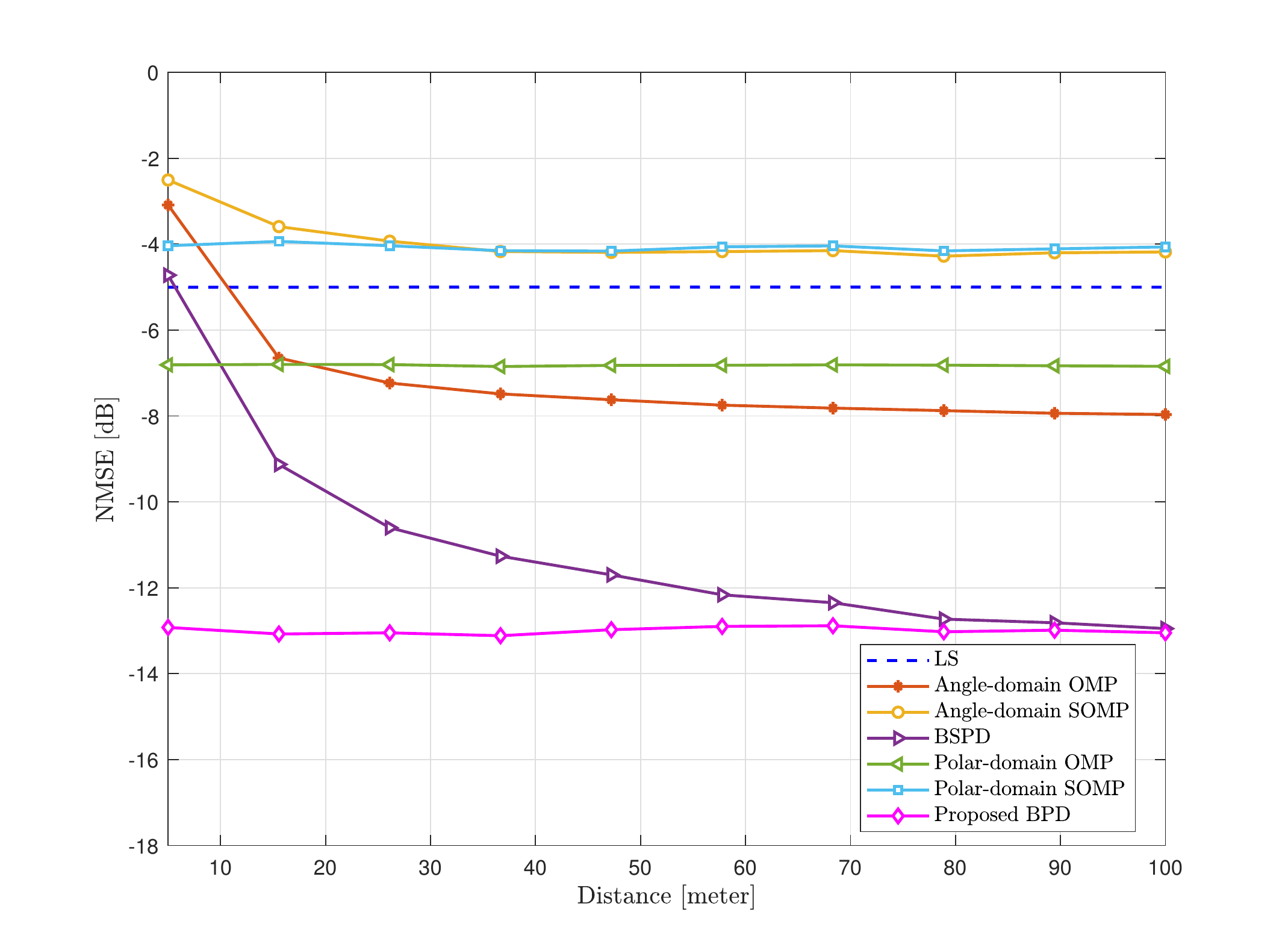}
  \vspace*{-1em}
  \caption{ NMSE performance against link distance. 
  }
  \label{fig:NMSE_R}
  \vspace*{-1em}
\end{figure}

First of all, the NMSE performance against distance is shown in Fig. \ref{fig:NMSE_R} to explain the influence of near-field effect. In this figure, we assume the distance $r_l$ is growing from $5$ meters to 100 meters, where we set $R_{\min} = R_{\max} = r_l$. The other parameter settings are as follows: 1) the $\text{SNR}$ is fixed as $5$ dB, 2) the bandwidth is $B = 10$ GHz, 3) and the pilot overhead is $P = 32$, with an observation dimension of $P N_{\text{RF}} = 128$. One can observe from Fig.  \ref{fig:NMSE_R} that the achieved NMSE of all far-field channel estimation methods (angle-domain OMP, SOMP, and BSPD) degrades with the decrease in distance. This is because these algorithms tailored for far-field communications neglect the impact of spherical wavefront. Besides, as the LS algorithm is able to work on all kinds of channels, its performance is robust to different distances. Similarly,  all near-field channel estimation algorithms also have a stable NMSE in both far-field and near-field environments by exploiting the polar-domain sparsity. Despite this distance-robust advantage, the achieved NMSE of the LS, polar-domain OMP, and SOMP methods are not satisfactory, because they fail to utilize the underlying beam split structure in wideband systems. Fortunately, the proposed BPD algorithm well captures the polar-domain frequency-dependent sparse support structure, and thus it outperforms all benchmark algorithms.

\begin{figure}
  \centering
  \includegraphics[width=3.5in]{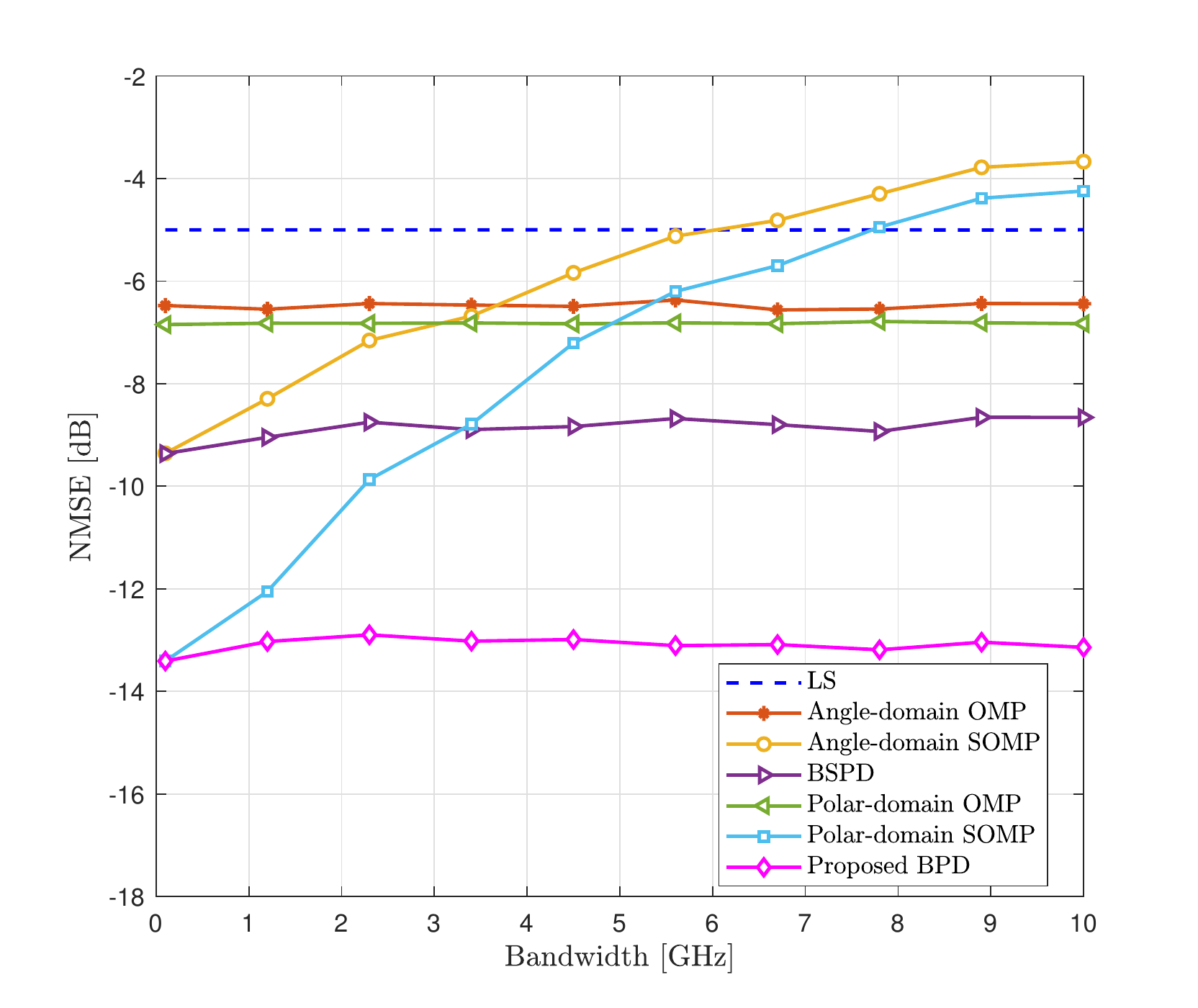}
  \vspace*{-1em}
  \caption{ NMSE performance against system bandwidth. 
  }
  \label{fig:NMSE_B}
  \vspace*{-1em}
\end{figure}
Then, Fig. \ref{fig:NMSE_B} illustrates the NMSE performance against system bandwidth, which is growing from $100$ MHz to 10 GHz. The other parameters settings are as follows: $\text{SNR} = 5$ dB, the pilot overhead is $P = 32$, $R_{\min} = 10$ meters, and $R_{\max} = 30$ meters. As the common sparse support set characteristic is not valid in wideband systems, the NMSE performance of angle-domain and polar-domain SOMP algorithms becomes worse and worse with the increase of bandwidth. In addition, the LS, angle-domain OMP, and polar-domain OMP schemes make no assumption on the wideband channel structure, where the channel of each subcarrier is independently estimated. As a consequence, these algorithms achieve a stable but non-satisfactory channel estimation accuracy for different bandwidth conditions. Moreover, the BSPD method accurately describes the angle-domain beam split pattern but ignores the distance-domain beam split pattern, so it is able to slightly improve the estimation accuracy for different bandwidth cases.  On the other hand, we can observe from Fig. \ref{fig:NMSE_B} that the proposed BPD algorithm realizes both stable and the most accurate estimation performance in all bandwidth cases. This is because the BPD scheme fully exploits the frequency-dependent sparse structure resulting from the near-field beam split effect.


\begin{figure}[h]
  \centering
  \includegraphics[width=3.5in]{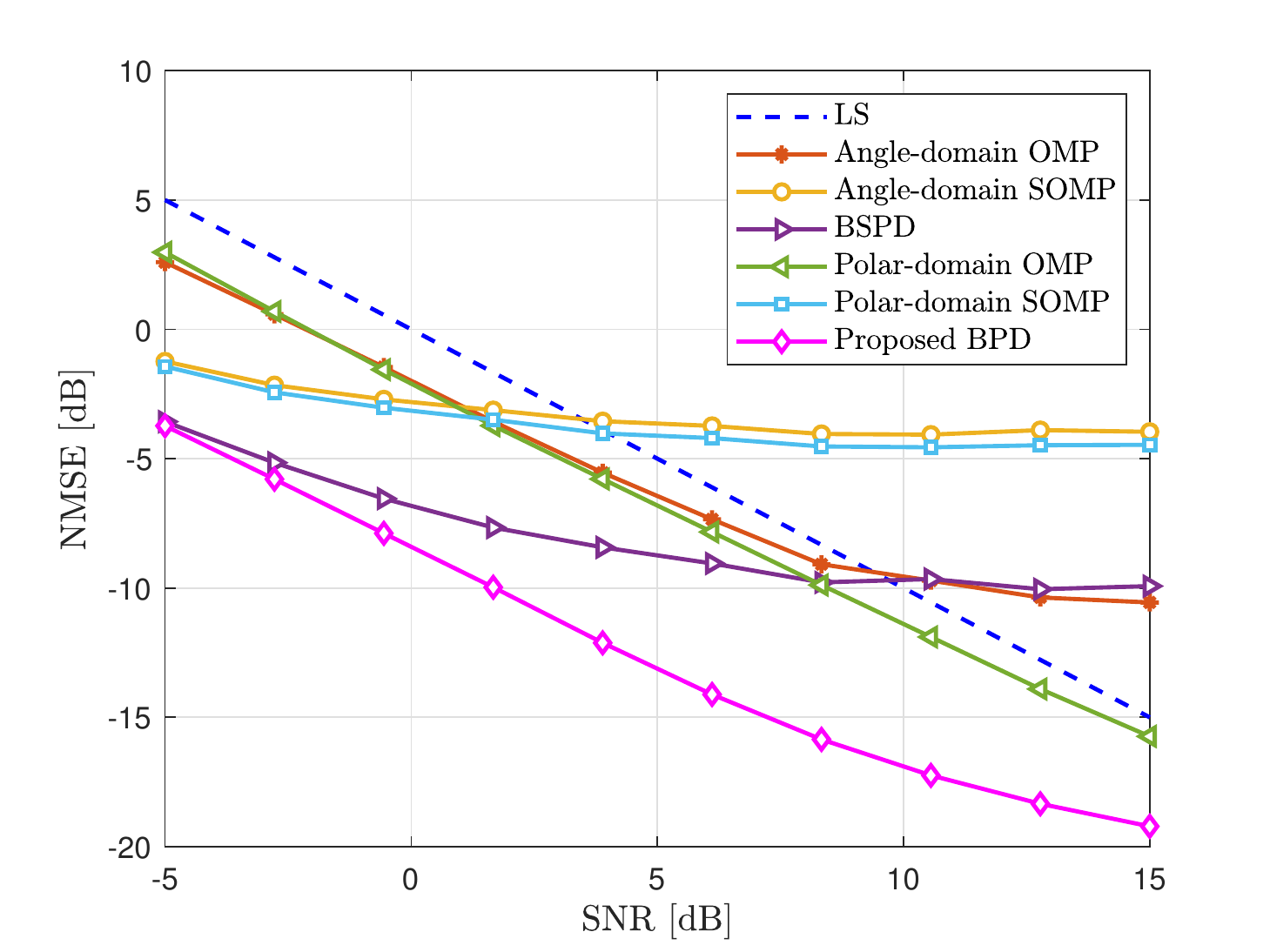}
  \vspace*{-1em}
  \caption{ NMSE performance against SNR. 
  }
  \label{fig:NMSE_SNR}
  \vspace*{-1em}
\end{figure}

In Fig. \ref{fig:NMSE_SNR}, we evaluate the achieved NMSE performance in different SNR conditions. Here, the SNR grows from -5 dB to 15 dB. The other parameter settings are set as below: $\text{B} = 10$ GHz, $R_{\min} = 10$ meters, $R_{\max} = 30$ meters, and $P = 32$. The accuracy achieved by all considered algorithms improves with the increase of SNR. It is clear that the proposed BPD method significantly outperforms all compared benchmarks, especially in  high SNR cases. For example, when the SNR is 7 dB, around 5 dB improvement of NMSE is realized compared to the BSPD method and the polar-domain OMP method. 

\begin{figure}[h]
  \centering
  \includegraphics[width=3.5in]{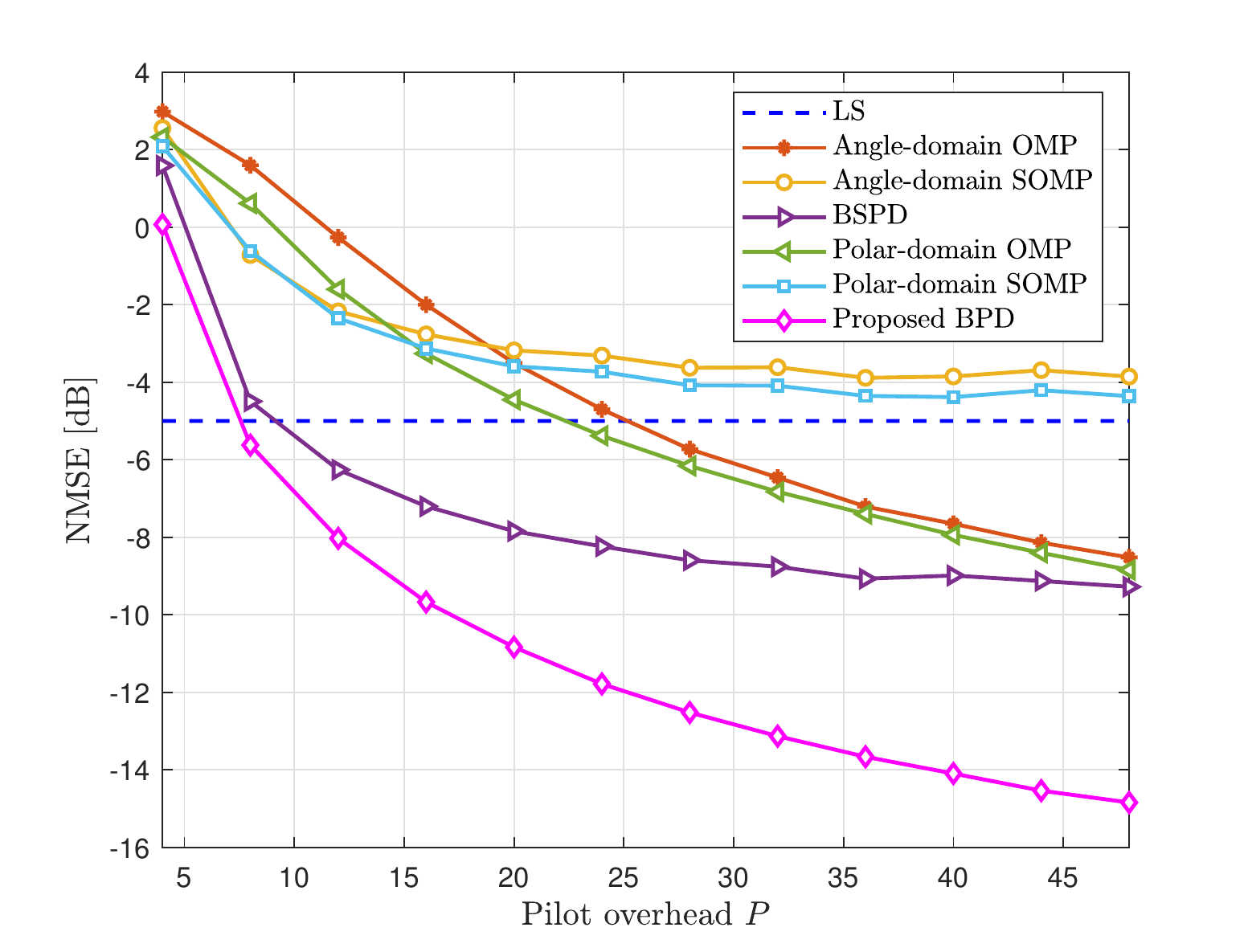}
  \vspace*{-1em}
  \caption{ NMSE performance against pilot overhead. 
  }
  \label{fig:NMSE_P}
  \vspace*{-1em}
\end{figure}
Eventually, in Fig. \ref{fig:NMSE_P}, the NMSE performance with respect to the pilot overhead $P$ is plotted. As illustrated in Fig. \ref{fig:NMSE_P}, the pilot overhead $P$ is increasing from 4 to 32, which corresponds to a compression ratio $\frac{N}{PN_{\text{RF}}}$ declining from $16$ to $2$. Then, the other parameters are set as: $\text{SNR} = 5$ dB, $R_{\min} = 10$ meters, $R_{\max} = 30$ meters, and $B = 10$ GHz. It is clear that the proposed BPD based method outperforms other existing channel estimation algorithms for all considered pilot lengths. This fact implies that the proposed method can be used to reduce the pilot overhead. Specifically, take the NMSE of -9 dB as a baseline, around 36 pilot overhead is required for the BSPD to achieve this NMSE baseline. On the contrary, only 12 pilot length is enough for the proposed BPD method to reach this baseline.  In this case, our scheme has the capability of reducing the pilot overhead by $66\%$.

  \section{Conclusions}\label{sec:5}
In this paper, we have investigated the channel estimation for wideband XL-MIMO communications in the presence of near-field beam split effect. 
Specifically, we first revealed the bilinear pattern of the near-field beam split effect, which indicated that the polar-domain sparse support set for each near-field channel path shows a linear structure over frequencies.  Then, we have proposed a BPD based channel estimation algorithm to recover each near-field channel path, by using the bilinear pattern to accumulate the largest polar-domain power from the entire bandwidth. Simulation results demonstrated our scheme is capable of achieving high channel estimation accuracy in all far-field/near-field/narrowband/wideband conditions. 
For future works, the bilinear pattern discovered in this paper can potentially be extended to tackle the relevant channel estimation issues in various near-field wideband communication scenarios, such as reconfigurable intelligent surface (RIS) communications~\cite{RISCE_Wei21} and cell-free massive MIMO communications~\cite{CF_Zhang21}.

\Acknowledgements{This work was supported in part by the National Key Research and Development Program of China (Grant No. 2020YFB1805005), in part by the National Natural Science Foundation of China (Grant No. 62031019), and in part by the European Commission through the H2020-MSCA-ITN META WIRELESS Research Project under Grant 956256.}



\end{document}